\documentclass{llncs}
\pdfoutput=1
\usepackage{graphicx}
\usepackage{multicol}
\usepackage{multirow}
\usepackage{amsmath}
\usepackage{url}

\begin{document}

\title{DNF Sampling for ProbLog Inference}
\author{Dimitar~Sht.~Shterionov, Angelika~Kimmig, Theofrastos~Mantadelis and Gerda~Janssens}
\institute{Department of Computer Science, Katholieke Universiteit Leuven\\
           \url{dimitar.shterionov@student.kuleuven.be}, \url{{firstname.lastname}@cs.kuleuven.be}}

\maketitle

\begin{abstract}
Inference in probabilistic logic languages such as ProbLog, an extension of Prolog with probabilistic facts, is often based on a reduction to a propositional formula in DNF. Calculating the probability of such a formula involves the disjoint-sum-problem, which is computationally hard. In this work we introduce a new approximation method for ProbLog inference which exploits the DNF to focus sampling. While this DNF sampling technique has been applied to a variety of tasks before, to the best of our knowledge it has not been used for inference in probabilistic logic systems. The paper also presents an experimental comparison with another sampling based inference method previously introduced for ProbLog.
\end{abstract}

\section{Introduction}
In the past few years, a multitude of formalisms combining probabilistic reasoning with logics, databases or logic programming has been developed, see for instance~\cite{DeRaedt08pilp,Getoor07} for overviews. To use such formalisms in statistical relational learning, efficient inference algorithms are crucial, as learning requires evaluating large numbers of queries. ProbLog~\cite{RAEDT07} is a simple extension of Prolog defining the success probability of a query in terms of random subprograms. Efficient inference algorithms for ProbLog have been implemented on top of the YAP-Prolog system~\cite{KIMMIG08}. ProbLog has been motivated by and applied to link mining in large collections of uncertain biological data, and its inference methods have been shown to increase the scalability of exact inference for probabilistic logic systems that do not rely on additional simplifying assumptions. However, as inference in such systems is computationally hard, approximation techniques are needed for complex queries. While~\cite{KIMMIG08} introduced a sampling based inference technique for ProbLog which directly exploits the distribution over subprograms defined by a ProbLog program, in this paper, we follow a more query-centered approach. We introduce \emph{DNF Sampling}, a new approximate inference technique for ProbLog based on the sampling scheme of~\cite{KARP83}, where in a first phase, as in ProbLog's exact inference, a Boolean formula in disjunctive normal form representing all proofs of the query is constructed. Samples are then drawn from this formula, thereby focussing on the subspace relevant for the current task. We experimentally compare both approaches in the context of biological networks, showing that Program Sampling has a better convergence than DNF Sampling.

The paper is organised as follows: We start by reviewing ProbLog and its key inference methods in Section~2. Section~3 introduces our new approximate inference method, and Section~4 reports on experiments comparing the different methods. After discussing related work in Section~5, we conclude in Section~6.

\section{ProbLog}
ProbLog is a probabilistic extension of Prolog inspired by typical machine learning applications. It is developed as a simple but powerful probabilistic logic programming language, and used for mining large biological networks (where nodes represent genes, proteins, and so on), with probability labels on their edges. As these tasks are computationally hard, the efficiency in processing complex queries is very important. For this reason, ProbLog is build on top of the state-of-the-art YAP-Prolog system. YAP is a high performance Prolog system, based on the Warren Abstract Machine (WAM) with different optimisations, which make it a suitable host for ProbLog.

ProbLog is closely related to other probabilistic logic systems such as PHA~\cite{POOLE93}, PRISM~\cite{SATO01}, and ICL~\cite{POOLE95}. However, PRISM and PHA impose additional assumptions to simplify probability calculation, and the ICL implementation ailog2 does not scale to larger problems. ProbLog's implementation is targeted at overcoming these limitations.

The syntax of a ProbLog program $T$ is similar to that of a Prolog one: it consists of facts and relations between them, but in the case of ProbLog a label is attached to some of the facts. That is, the program can be split into a set of labelled facts, where each $p_i :: f_i$  defines a fact $f_i$ with probability of occurrence $p_i$, and a Prolog program using those facts, which encodes background knowledge ($BK$). We denote the set of all $f_i$ (without probability label) by $L_T$. Probabilistic facts correspond to mutually independent random variables (RVs), which together define a probability distribution over all ground logic programs $L \subseteq L_T$:
\begin{equation}
    P(L|T)=\prod\nolimits_{f_i\in L}p_i\prod\nolimits_{f_i\in L_T\backslash L}(1-p_i)\ldotp
\label{eq:subprog}
\end{equation}
We use the term \emph{possible world} to denote the least Herbrand model of such a subprogram $L$ together with the background knowledge $BK$ and, by slight abuse of notation, use $L$ to refer to both the set of sampled facts and the corresponding world.

\begin{figure}
  \begin{tabular}{cl}
\multirow{7}{*}{\includegraphics[scale=0.40]{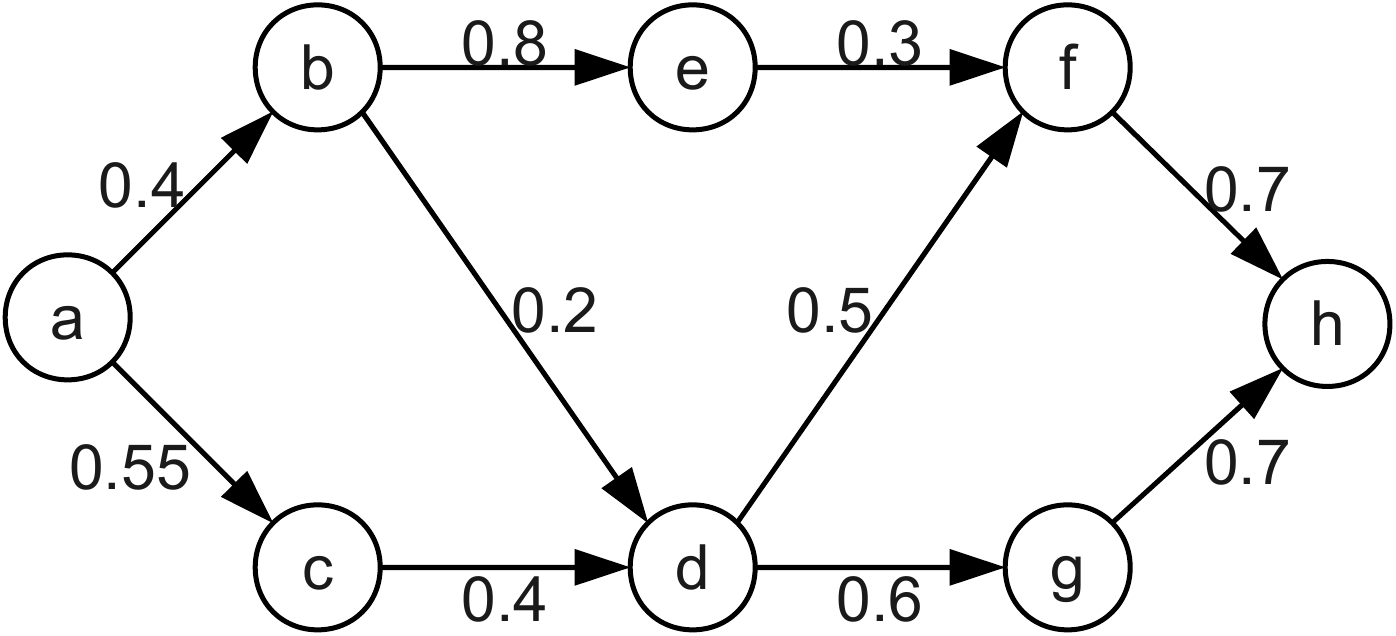}} &\verb|0.40 :: edge(a,b).  0.55 :: edge(a,c).|\\
&\verb|0.80 :: edge(b,e).  0.20 :: edge(b,d).|\\
&\verb|0.40 :: edge(c,d).  0.30 :: edge(e,f).|\\
&\verb|0.50 :: edge(d,f).  0.60 :: edge(d,g).|\\
&\verb|0.70 :: edge(f,h).  0.70 :: edge(g,h).|\\
&\verb|path(X, Y) :- edge(X, Y).|\\
&\verb|path(X, Y) :- edge(X, Z), path(Z, Y).|\\
&\\
\textbf{(a)} Probabilistic graph & \multicolumn{1}{c}{\textbf{(b)} ProbLog program}
  \end{tabular}

  \caption{An example of a probabilistic graph and the corresponding ProbLog program.}
  \label{fig:1}
\end{figure}

Figure~\ref{fig:1} shows a typical example of a probabilistic graph encoded in ProbLog. One can query the probability that a path exists between two nodes in the graph. As it can be noticed from the graph of Figure~\ref{fig:1}, there are several possible paths between two nodes. For example between nodes $b$ and $f$, we have two possible paths: $b \rightarrow e \rightarrow f$ and $b \rightarrow d \rightarrow f$. In ProbLog, querying for the probability of \verb|path(b, f)| means asking for the probability that a randomly selected subgraph contains a path from $b$ to $f$. Such subgraphs can contain the edges of the path $b \rightarrow e \rightarrow f$ or those of the path $b \rightarrow d \rightarrow f$, but also all of them or even many more. The success probability $P_s(q|T)$ of a query $q$ can now be defined as follows:
\begin{equation}
  P_s(q|T) = \sum_{L \subseteq L_T}{P(q|L) \cdot P(L|T)}
\label{eq:success}
\end{equation}
where $P(q|L)$ is $1$ if there is a substitution $\theta$ such that $q\theta$ is entailed by the union of $L$ and the background knowledge $(L \cup BK \models q\theta)$, and $0$ otherwise. Equation~(\ref{eq:success}) states that the success probability of the query \verb|path(b, f)| can be calculated by summing the probabilities of all subgraphs which include at least one path connecting nodes $b$ and $f$. As the number of subprograms to be considered is exponential in the number of probabilistic facts, this approach quickly becomes infeasible with increasing problem size. The ProbLog system therefore uses a different approach, which will be discussed in Section~\ref{sec:exact}.

A second inference task in ProbLog is the identification of the \emph{best proof} or explanation of a query.
An explanation, also called proof here, is a set of probabilistic facts $\alpha \in L_T$ which satisfies the following properties~\cite{BRACHMAN04}:
\begin{enumerate}
  \item it is sufficient to account for $q$, i.e. $BK \cup \alpha \models q$,
  \item it is not ruled out by the $BK$, i.e. $BK \cup \alpha$ is consistent,
  \item there is no $\beta \subset \alpha$ such that $1$ and $2$ hold for $\beta$.
\end{enumerate}

In Equation~(\ref{eq:success}), the sum goes over those subprograms that contain \emph{some} proof of the query. When considering a specific explanation, this is further restricted to those subprograms containing all facts of that explanation. Therefore the probability of an explanation $\alpha$ is given by the following formula:
\begin{equation}
  P(\alpha|T) = \sum_{\alpha \subseteq L \subseteq L_T}{P(L|T) = \prod_{f_i \in \alpha}{p_i}}.
\label{eq:expl}
\end{equation}
As there may exist many explanations $\alpha$ for a  query $q$, the one with the highest probability is used to define the \emph{explanation probability} of $q$:
\begin{equation}
P_x(q|T)  =  \max\nolimits_{\alpha\in E(q)}P(\alpha|T)
=   \max\nolimits_{\alpha \in E(q)} \prod_{c_i \in \alpha}p_i,\label{eq:p_exp}
\end{equation}
where $E(q)$ is the set of all explanations for query $q$.

For our example, both probabilities as given in Equations (\ref{eq:success}) and (\ref{eq:p_exp}) are easily computed even by hand: the success probability is $P_s(path(b, f)|T1) = 0.316$ (note that it is sufficient to consider the graph restricted to nodes $b$, $e$, $d$ and $f$ when listing subprograms for this query), and the explanation probability is $\max(0.8\cdot 0.3, 0.2\cdot 0.5) = \max(0.24, 0.1) = 0.24$, but for complex problems this could consume large amounts of time and memory. ProbLog therefore follows different strategies to obtain success probabilities, which we will briefly discuss next.

\subsection{Exact Inference} 
\label{sec:exact}
As iterating over possible subprograms as done in Equation~(\ref{eq:success}) is infeasible for most programs, ProbLog's exact inference instead employs a reduction to propositional formula in disjunctive normal form (DNF). As stated earlier, probabilistic facts can be seen as RVs, implying that a proof can be represented as a conjunction of such facts. The set of all proofs can then be represented as a disjunction, producing a DNF formula. The success probability then corresponds to the probability of this formula being true. In our example we obtain the formula $(e(b,e) \wedge e(e,f)) \vee (e(b,d) \wedge e(d,f))$ where $e/2$ denotes edge. Each proof's probability is calculated as the product of the probabilities of its facts, cf.~Equation~(\ref{eq:expl}). Following the simple logic of conjunction and disjunction we could infer that the summation of all proofs' probabilities will produce the final result. However, this is only true under specific conditions, namely if each possible world permits at most one proof of the query. PRISM requires that programs respect these conditions, which means that proofs have to be \emph{mutually exclusive} (w.r.t.~occurrence in possible worlds). In our example, these conditions are not met:  we would obtain $0.34$, while the correct value is $0.316$. One way to deal with this problem is to consider the conjunctions in the DNF sequentially, and to replace each proof or conjunction $\alpha_i$ by its conjunction with the negation of all the proofs after it, that is, by $\alpha_i \wedge \bigwedge_{j\geq i}\neg\alpha_j$. In this way, each possible world permits at most one such extended proof. Note however that the resulting formula needs further manipulation to be transformed into a sum of products which can be used for easy calculation. For the previous example this will produce:
\begin{align*}
  P_s(path(b,f)|T) = & P((e(b,e) \wedge e(e,f)) \vee (e(b,d) \wedge e(d,f))|T)\\
                   = & P((e(b,e) \wedge e(e,f)) \wedge \neg(e(b,d) \wedge e(d,f))) + P(e(b,d) \wedge e(d,f))\\
                   = &0.8 \cdot 0.3 \cdot (1-(0.2 \cdot 0.5))+ 0.2 \cdot 0.5 = 0.316.
\end{align*}

Unfortunately this type of technique is feasible only for small formulae. This problem is known as the disjoint-sum-problem (as it is concerned with making the contributions of the different parts of the summation non-overlapping) and is \#P-complete~\cite{VALIANT79}. While both PRISM and PHA avoid the problem by imposing the requirement of mutually exclusive proofs, ICL uses a symbolic disjoining technique to refine proofs into formulae describing mutually exclusive sets of possible worlds~\cite{POOLE95}. The ProbLog system deals with it using Reduced Ordered Binary Decision Diagrams (BDDs), which are graphical representations of a Boolean function over a set of variables, which significantly extends scalability of inference. Still, at some point one needs to resort to approximate inference techniques. Exact Inference bottlenecks lie in two separate steps, a first possible overhead both for space and time is collecting the proofs which can be exponential in the number of probabilistic facts. Even when one successfully collects all the proofs, solving the disjoint-sum-problem is a \#P-complete problem and the BDD approach easily can explode in space.

\subsection{Approximate Inference: Program Sampling} 
An alternative approach to inference is the use of Monte Carlo methods, that is, to use the ProbLog program to generate large numbers of random subprograms and to use those to estimate the probability. More specifically, such a method proceeds by repeating the following steps:
\begin{enumerate}
\item sample a logic (sub)program $L$ from the ProbLog program
\item search for a proof of the initially stated query $q$ in the sample $L\cup BK$
\item estimate the success probability as the fraction $P$ of samples which hold a proof of the query
\end{enumerate}
The implementation of this approach for ProbLog, as described in~\cite{KIMMIG08}, takes advantage of the independence of probabilistic facts to generate samples lazily while proving the query, that is, sampling and searching for proofs are interleaved. 
To assess the precision of the current estimate $P$, at each $m$ samples the width $\delta$ of the $95\%$ confidence interval is approximated as
\begin{equation}
 \delta = 2 \cdot \sqrt{\frac{P \cdot (1-P)}{N}}
\end{equation}
If the number of samples $N$ is large enough the interval of confidence becomes smaller, and the certainty that the estimate is close to the true probability of the query increases. We will refer to this method as \emph{Program Sampling} here to avoid confusion with the method that will be introduced in Section~\ref{sec:dnf_sampling}.

\section{DNF Sampling}\label{sec:dnf_sampling}
Program Sampling generates samples by exploring the SLD tree, which can be expensive if there are many failing derivations. In this section, we therefore introduce a new sampling based method for ProbLog which focuses sampling on possible worlds containing a proof of the query of interest. This method first constructs the DNF for the query as in exact inference, and then applies the Monte-Carlo algorithm of Karp and Luby~\cite{KARP83} to estimate the probability of the DNF. In the following, we will discuss the algorithm and its implementation for ProbLog in more detail.

\subsection{An Example}
Let us start by considering an example, namely the DNF $F=(a\wedge b\wedge c) \vee (b\wedge c\wedge d) \vee (b\wedge d\wedge e)$, where we assume a probability of $0.5$ for each random variable. The probability of a conjunction is easily calculated as the product of the probabilities of the involved facts, cf.~Equation (\ref{eq:expl}); in this case, each of the conjunctions in the DNF thus has probability $P(c_i)=0.125$. As there are five random variables, each possible world has probability $0.03125$.

Figure~\ref{fig:boxes} shows the possible worlds associated to each of the three conjunctions. Summing the probabilities of the conjunctions, we obtain $S(F) = 0.375$. However, as can be seen in Figure~\ref{fig:boxes}, the conjunctions are not mutually exclusive: all of them are true in world $(1)$, and two of them are true in worlds $(2)$ and $(5)$. In total, there are $8$ different worlds to be taken into account for the probability of $F$, which therefore is only $P(F)=0.25$. 

DNF Sampling now associates each possible world $w$ to the first conjunction that is true in $w$, that is, worlds (1) and (2) are associated to $a\wedge b\wedge c$, world (5) to $b\wedge c\wedge d$. Samples are generated by first sampling a conjunction $c_i$ with probability $P(c_i)/S(F)$, and then generating a possible world by setting the truth values of the variables in $c_i$ such that $c_i$ is true, and sampling truth values for remaining variables. For instance, we could obtain $b\wedge c\wedge d$ and world (2) in this way, but as world (2) is associated to $a\wedge b\wedge c$, this sample would be considered negative, whereas $a\wedge b\wedge c$ together with world (2) would be considered positive. Informally speaking, the sampling procedure thus rejects half of the contribution of world (2), thereby reducing it to its true value.
As for each pair $(c_i,w)$, the probability of sampling that pair is $P(w)/S(F)$ and thus proportional to $P(w)$, given a sufficient number $N$ of samples, the fraction $N_{accepted}/N$ of positive samples approaches $P(F)/S(F)$, and we can thus estimate $P(F)$ as $ S(F)\cdot N_{accepted}/N$.

\begin{figure}[t]
\centering
\begin{tabular}{|llllll|}
\hline
(1) & a & b & c & d & e \\\hline
(2) & a & b & c & d &   \\\hline
(3) & a & b & c &   & e \\\hline
(4) & a & b & c &   &   \\\hline
\end{tabular}
\hspace{8mm}
\begin{tabular}{|llllll|}
\hline
(1) &a & b & c & d & e \\\hline
(2) &a & b & c & d &   \\\hline
(5) &  & b & c & d & e \\\hline
(6) &  & b & c & d &   \\\hline
\end{tabular}
\hspace{8mm}
\begin{tabular}{|llllll|}
\hline
(1) & a & b & c & d & e \\\hline
(7) & a & b &   & d & e \\\hline
(5) &   & b & c & d & e \\\hline
(8) &   & b &   & d & e \\\hline
\end{tabular}
\caption{Possible worlds associated to the conjunctions of $(a\wedge b\wedge c) \vee (b\wedge c\wedge d) \vee (b\wedge d\wedge e)$.}
\label{fig:boxes}
\end{figure}

\subsection{Algorithm}
We now formalise the algorithm. Let $F = C_1\vee\ldots\vee C_m$ be a propositional DNF for query $q$ in the ProbLog program $T$, where the $C_i$ contain neither contradictions nor multiple occurrences of the same variable. We denote possible worlds -- truth value assignments to all random variables  -- by $w$. The space from which samples are drawn is defined as $U=\{(w,i) | w\models C_i\}$, and we associate each possible world $w$ to the first conjunction that is true in $w$, that is, the samples that will be accepted are those from $A=\{(w,i) | w\models C_i \wedge\forall j < i:  w\not\models C_j\}$. For each possible world $w$ with $w\models F$, $U$ thus contains a pair $(w,i)$ for each $C_i$ that is true in $w$, whereas $A$ only contains the pair with minimal $i$. We define the sum of probabilities for DNF $F$ as
\begin{equation}
  S_T(F) = \sum_{i=0}^{m}\prod_{f_j\in C_i}p_j
\label{equ:2}
\end{equation}
Note that if conjunctions are mutually exclusive as discussed in Section~\ref{sec:exact}, $S_T(F)$ is equal to the probability of $F$ being true, but it can be much higher in general.

DNF Sampling generates $N$ samples in the following way
\begin{enumerate}
\item randomly choose $C_i$ according to $P(C_i|T)/S_T(F)$
\item randomly choose a possible world $w$ where $C_i$ is true
\item increment $N_{accepted}$ if $(w,i)\in A$
\end{enumerate}
The probability of formula $F$ is then estimated as
\begin{equation}
P_{DNF}(q|T) = S_T(F) \cdot \frac{N_{accepted}}{N}
\label{eq:4}
\end{equation}
Note that depending on the structure of the problem and the value of $S_T(F)$, estimates based on small numbers of samples may not be probabilities yet, that is, be larger than one, especially if the actual probability is close to one. This is due to the fact that a sufficient number of samples is needed to identify overlap between conjunctions by means of sampling and to accordingly scale down the overestimate $S_T(F)$.

\subsection{Convergence}
DNF Sampling is an instance of the fully polynomial approximation scheme of Karb and Luby~\cite{KARP83}, that is, the number of samples required for a given level of certainty is polynomial in the input length (the DNF in our case). For formal detail, we refer to~\cite{KARP83}, and instead give a rough illustration here. The algorithm uses the normalization factor $S_T(F)$ from equation~(\ref{equ:2}), therefore, in each sampling step, the probability that the $i^{th}$ conjunction is sampled is $P(C_i) / S_T(F)$. It is then completed into a possible world according to the fact probabilities.
This possible world will be accepted exactly if all conjunctions with smaller index are false in it. So the probability of sampling a specific $C_i$ and a world which will be accepted for this conjunction is $P(C_i \wedge \neg DNF_{i-1}) / S_T(F)$, where $DNF_i = \bigvee_{j=1\ldots i}C_j$. As each world can only be accepted for exactly one conjunction, the probability of sampling an arbitrary world that will be accepted is the sum of this probability over all conjunctions, and for $N$ samples, the estimated number of accepted possible worlds and the corresponding probability estimate thus are:
\begin{align*}
E[N_{accepted}] = & N \cdot \frac{P(C_1) + P(C_2 \wedge \neg DNF_{1}) + ... + P(C_n \wedge \neg DNF_{n-1})}{S_T(F)} \\
P_{estimated} = & \frac{E[N_{accepted}]}{N} \cdot S_T(F) \\
              = & P(C_1) + P(C_2 \wedge \neg DNF_{1}) + ... + P(C_n \wedge \neg DNF_{n-1})
\end{align*}
The last line corresponds to one way of solving the disjoint-sum-problem for the original DNF, which is exactly what the purpose of the algorithm is.

In the current implementation of DNF Sampling, we use the same stopping criterion as for Program Sampling. Investigating alternative criteria tailored towards the new method and its convergence properties as analyzed in~\cite{KARP83} is part of future work.

\subsection{Implementation}
Here we explain some details about the ProbLog implementation of the algorithm. We will assume that the DNF is stored as a doubly linked list of proofs, which is sequentially accessible and can be traversed in both directions.\footnote{ProbLog uses a trie datastructure for this purpose, but as this is not exploited in the implementation, we simplify for ease of presentation.} The algorithm starts from the start of the list and traverses it in forward direction. For each proof, it calculates its probability as the product of its facts' probabilities, cf.~Equation~(\ref{eq:expl}), as well as the sum of the probabilities of all proofs processed so far, cf.~Equation~(\ref{equ:2}). The sum associated with the last proof is the normalisation factor $S_T(F)$ used in sampling.

Each sample first determines a  proof $\alpha_i$ (clause $C_i$). To this aim, we sample a threshold $T \in [0, 1)$ uniformly at random and use binary search on the memoized summed probabilities to identify $\alpha_i$ as the first proof whose associated sum exceeds $T \cdot S_T(F)$.

To exemplify this procedure, let us consider the graph from Figure~\ref{fig:1} and the list holding the proofs for the query \verb|path(a, h)| as shown in Figure~\ref{figure:2}, where initial computations are already included. Assuming threshold $T = 0.4$ has been chosen, the algorithm determines the $3^{rd}$~proof to be the chosen proof $\alpha_i$. 

\begin{figure}
\centering
 \begin{tabular}{llll}
\# & Proof & Probability      & Sum \\\hline
1. & [a, b, e, f, h] & 0.0672 & 0.0672 \\
2. & [a, b, d, f, h] & 0.0280 & 0.0952 \\
3. & [a, b, d, g, h] & 0.0336 & 0.1288 \\
4. & [a, c, d, g, h] & 0.0924 & 0.2212 \\
5. & [a, c, d, f, h] & 0.0770 & 0.2982 \\\hline
\multicolumn{3}{l}{Normalizing factor $S_T(F)$:} & 0.2982
 \end{tabular}
\caption{A list holding the proofs of the query \texttt{path(a, h)} and their probabilities.}
\label{figure:2}
\end{figure}

We now need to extend the sample by a possible world where $\alpha_i$ is true. As in Program Sampling, a lazy strategy that exploits independence of random variables is followed for this purpose, where sampling the world is interleaved with determining whether the sample will be accepted. First, we set the truth values of all variables in $\alpha_i$ such that $\alpha_i$ is true. Truth values are recorded in an array\footnote{We used an array to ensure fast look-up time.} as either true, false, or undetermined.
Next, the algorithm traverses the list backwards starting from $\alpha_i$. For each proof $\alpha_j$, we check the truth value of each of its variables, and if it is not yet fixed in the current world, determine it by sampling. As soon as $\alpha_j$ is determined to be false in the current world, that is, the known truth value of a variable is the opposite of the one needed to make the current proof true, we stop its evaluation and continue with the next proof $\alpha_{j-1}$. If a proof $\alpha_j$ with $j<i$ turns out to be true in the current world, we know that the sample will not be accepted, and the algorithm terminates the current iteration. If the last proof has been reached and determined to be false in the current world, the sample is accepted and the counter $N_{accepted}$ is incremented. In both cases, the next iteration is started, until the desired number $N$ of samples is reached. The final step of the algorithm includes the calculation of the approximated probability as shown in Equation~(\ref{eq:4}).

\section{Experiments}

In this section, we experimentally evaluate our new algorithm. The purpose of these initial experiments is:
\begin{enumerate}
  \item to show that DNF Sampling converges to the exact probability when the proofs are mutually exclusive,
  \item to compare convergence of DNF Sampling with Program Sampling, and
  \item to experimentally evaluate the performance of DNF Sampling.
\end{enumerate}
To this end we used three different benchmarks.

\begin{figure}
  \includegraphics[trim = 1.2cm 1.2cm 1.2cm 1.2cm, clip, width=12cm]{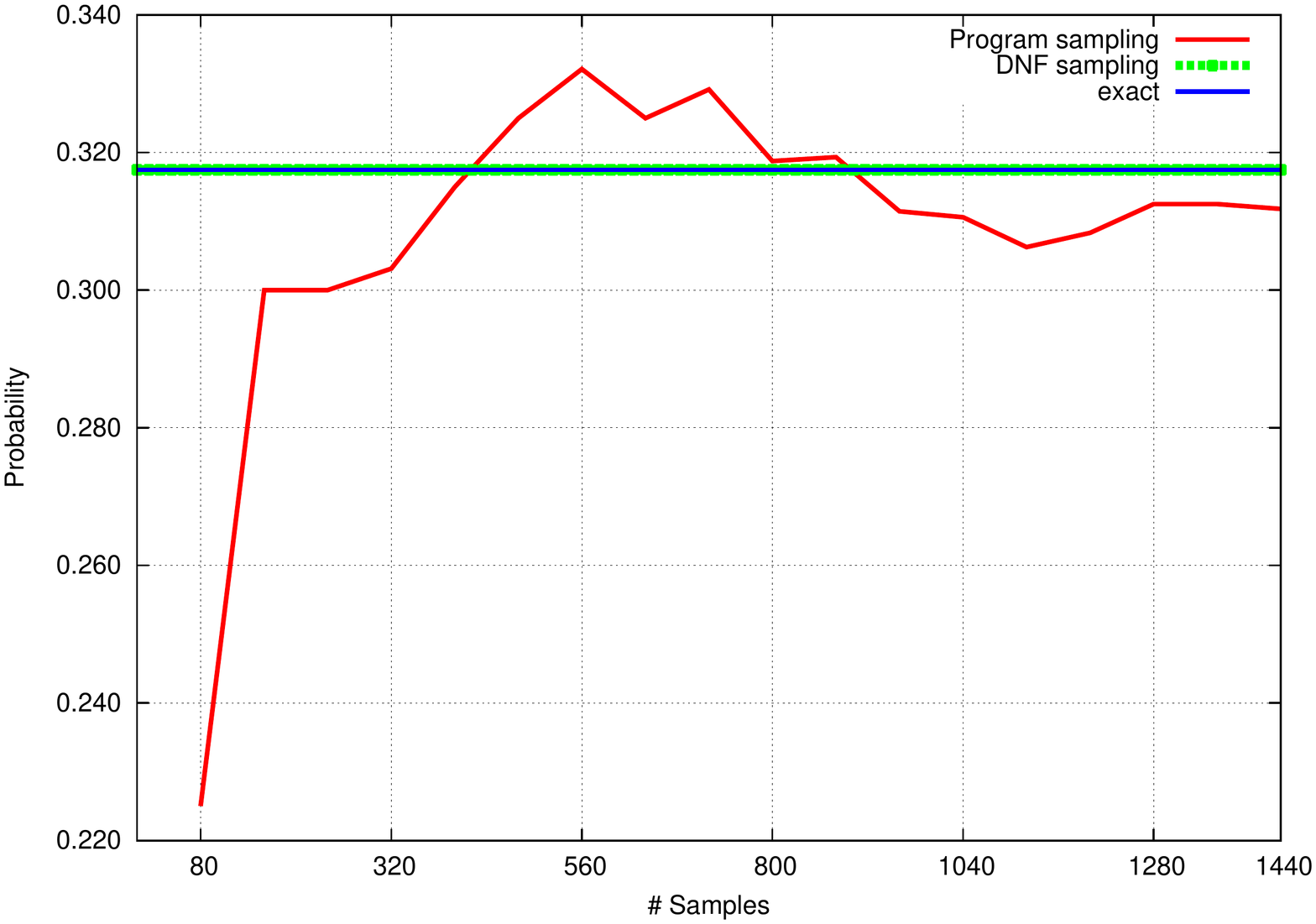}
  \caption{DNF Sampling convergence vs. Program Sampling convergence vs. Exact, for the case of mutually exclusive proofs with query \texttt{bloodtype(a,g4\_f144)}.}
  \label{fig:bloodtype}
\end{figure}

The first benchmark is a Bayesian Network that encodes a family tree of $15$ generations, where queries ask for the bloodtype of members of this pedigree. Here, proofs are always mutually exclusive, which means that each sample will be accepted, as no previous proof can be true in the same world.  DNF Sampling therefore converges immediately to the exact probability, as confirmed by the results  shown in Figure~\ref{fig:bloodtype}. These results also illustrate that focussing sampling on the proofs can improve convergence compared to Program Sampling, where the entire space of possible worlds needs to be explored to reach convergence.

For the following benchmarks, we estimate probabilities after each thousand samples and report averages over ten runs, as well as their standard deviation as error bars.

\begin{figure}
  \includegraphics[trim = 1.2cm 1.2cm 1.2cm 1.2cm, clip, width=12cm]{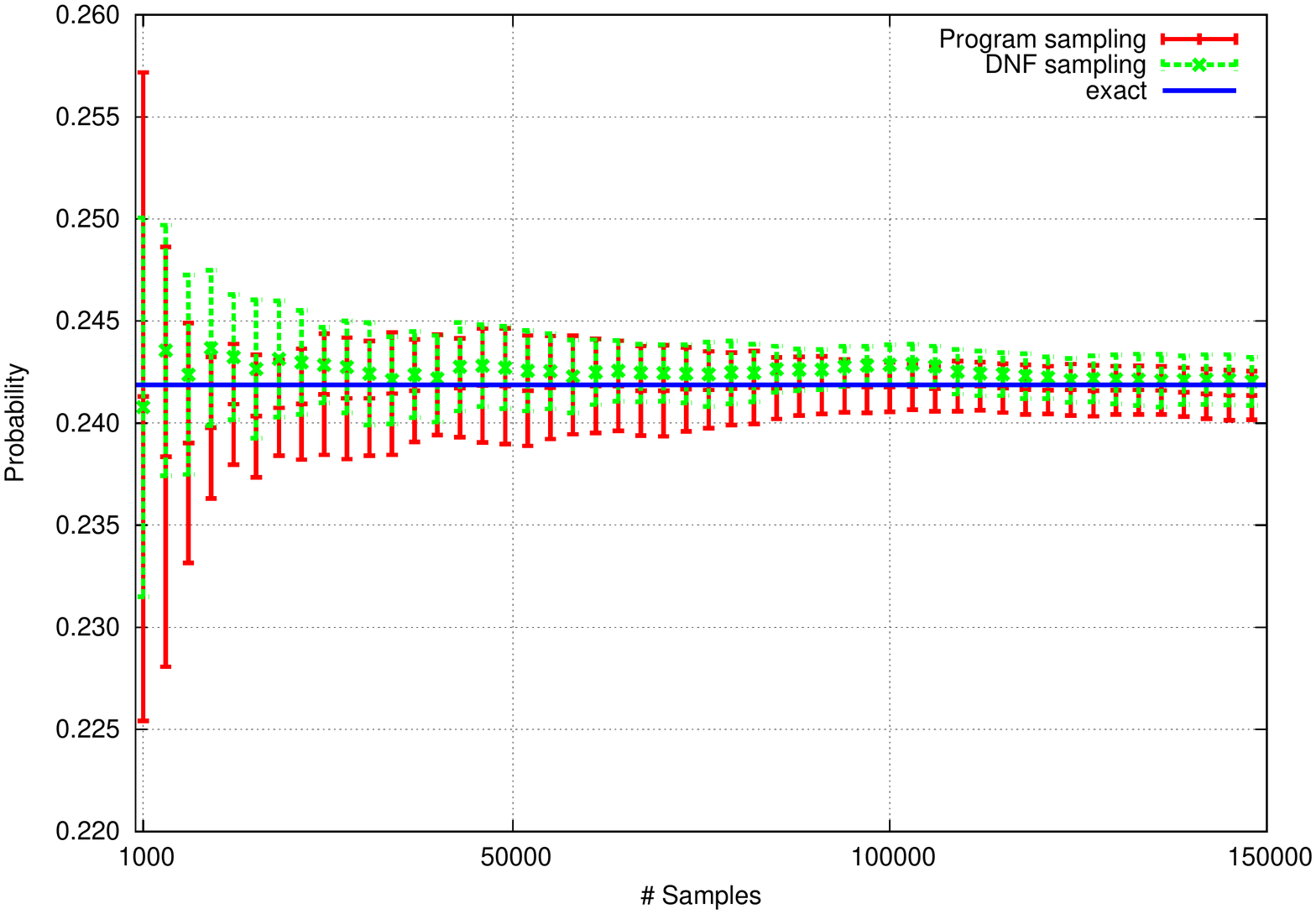}
  \caption{DNF Sampling convergence vs. Program Sampling convergence vs. Exact, for: \texttt{path('HGNC\_983','HGNC\_620',10)}.}
  \label{fig:alzw3_10}
\end{figure}

\begin{figure}
  \includegraphics[trim = 1.2cm 1.2cm 1.2cm 1.2cm, clip, width=12cm]{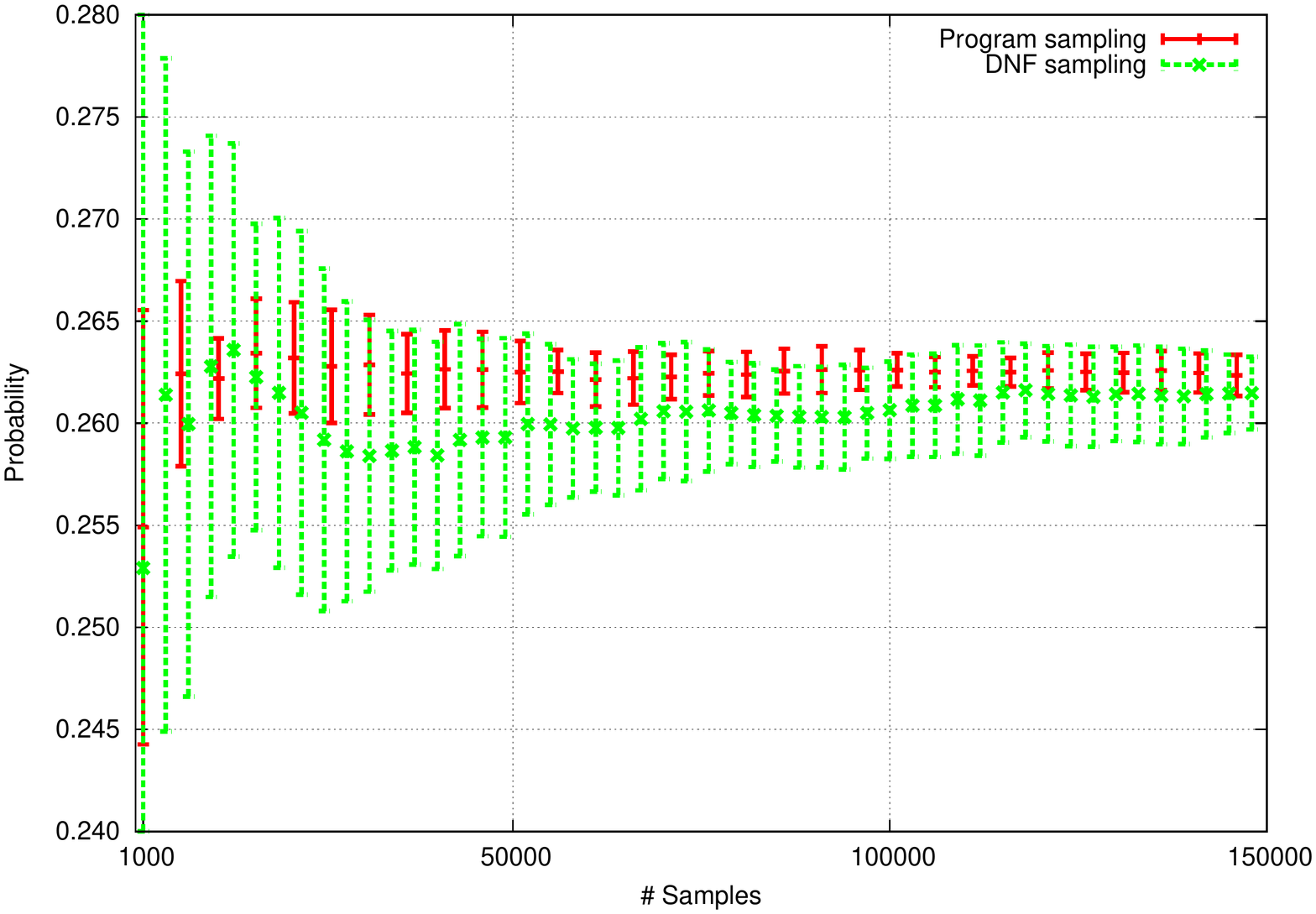}
  \caption{DNF Sampling convergence vs. Program Sampling convergence, for: \texttt{path('HGNC\_983','HGNC\_620',12)}.}
  \label{fig:alzw3_12}
\end{figure}

The second benchmark is the medium size Alzheimer graph of~\cite{KIMMIG08} with a \verb|path/3| predicate that defines paths between nodes with a maximal number of edges given by the third argument~$N$. We consider the same pair of nodes \verb|('HGNC_983', 'HGNC_620')| for $N=10$ and $N=12$, corresponding to $167$ and $2120$ proofs respectively. The exact method has difficulty in evaluating the DNF for paths between those two nodes for~$N > 10$.
Note that if the number of samples is smaller than the number of proofs, the algorithm cannot use all proofs for its estimate. The results are summarised in Figures~\ref{fig:alzw3_10}~and~\ref{fig:alzw3_12}.

Figure~\ref{fig:alzw3_10} presents the results for query \verb|path('HGNC_983','HGNC_620',10)|,  where the exact probability can still be calculated. The results confirm that both approximation methods converge towards the exact probability. Furthermore, we observe that Program Sampling converges faster. 

Figure~\ref{fig:alzw3_12} shows the evaluation of a computationally harder query, for which the exact value can not be calculated. Again we see that the sampling methods tend to converge towards one value. And especially after the point of $50,000$ samples, 99.96\% of the results are within an interval of size $0.005$. Reaching towards the $100,000^{th}$ sample the convergence in one value is obvious. The specificity of different problems will require different number of samples for convergence, but we show empirically that the method will converge towards the exact success probability.

Our third benchmark uses the same Alzheimer data set, but with nodes \verb|('HGNC_582', 'HGNC_620')| for $N=6$ and $N=8$, collecting $145$ and $1836$ proofs respectively. The exact method has difficulty in processing the DNF for~$N > 6$. These queries have the characteristic that the propability $P \rightarrow 1.0$. We present this benchmark as it is a worst case scenario for DNF Sampling.

\begin{figure}
  \includegraphics[trim = 1.2cm 1.2cm 1.2cm 1.2cm, clip, width=12cm]{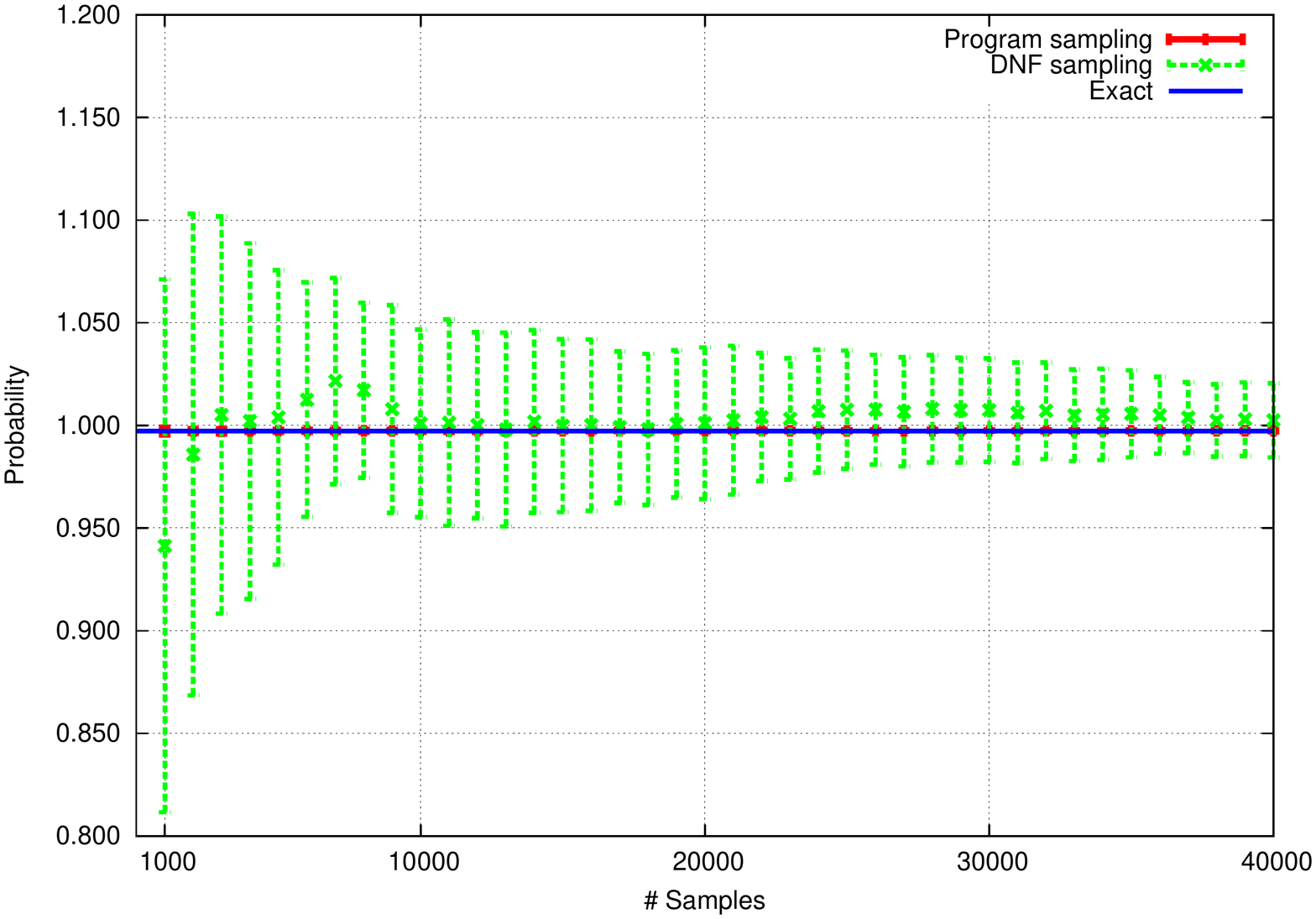}
  \caption{DNF Sampling convergence vs. Program Sampling convergence vs. Exact, for: \texttt{path('HGNC\_582','HGNC\_620',6)}.}
  \label{fig:alzw3_6}
\end{figure}

\begin{figure}
  \includegraphics[trim = 1.2cm 1.2cm 1.2cm 1.2cm, clip, width=12cm]{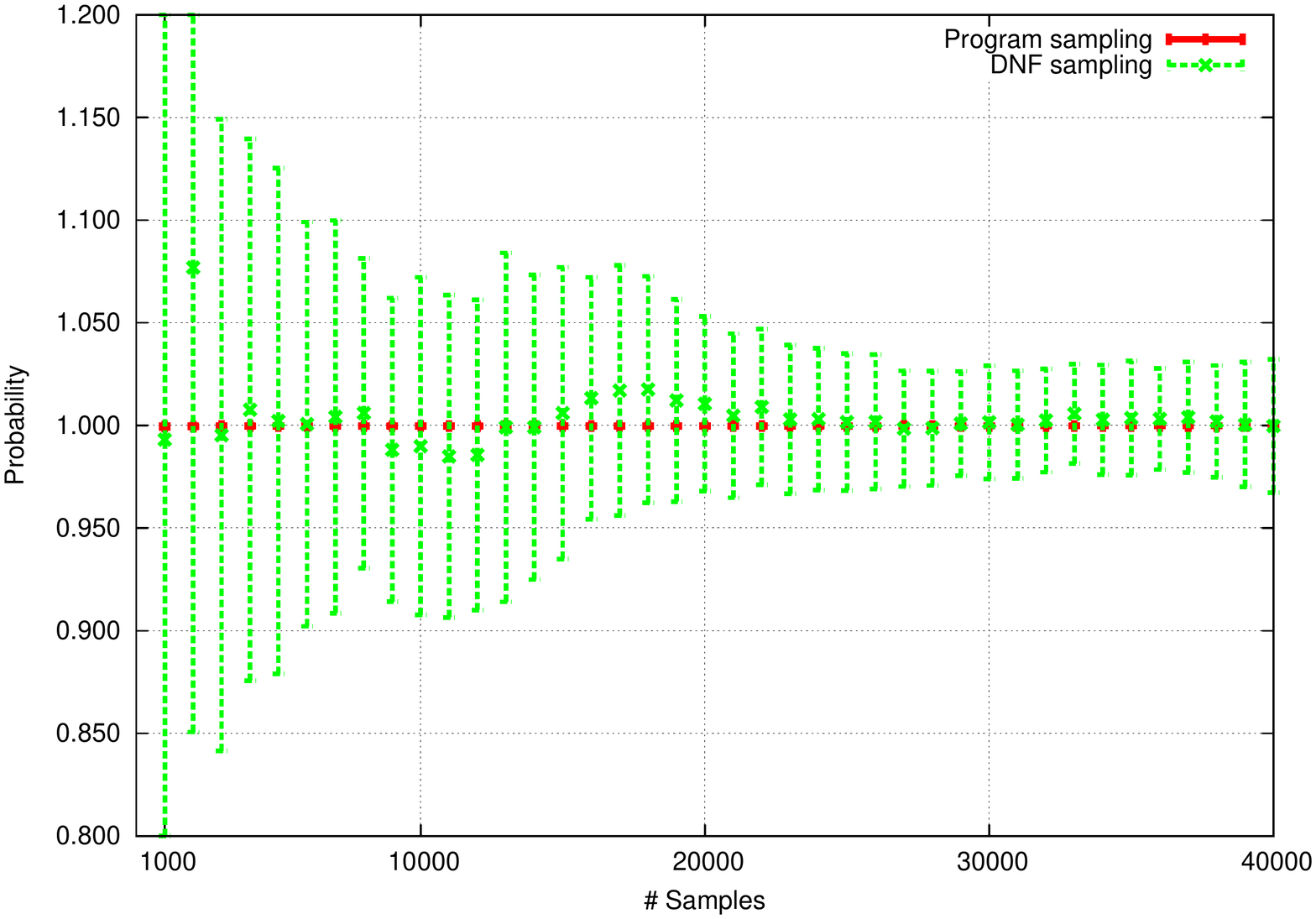}
  \caption{DNF Sampling convergence vs. Program Sampling convergence, for: \texttt{path('HGNC\_582','HGNC\_620',8)}.}
  \label{fig:alzw3_8}
\end{figure}

One can notice in Figures~\ref{fig:alzw3_6}~and~\ref{fig:alzw3_8}, that DNF Sampling can result in estimating a probability higher than $1.0$, this is not an erroneous result as one can infer from Equation~(\ref{eq:4}). One can also notice that DNF Sampling has difficulty to converge at the second query of Figure~\ref{fig:alzw3_8}.

\begin{figure}
  \includegraphics[width=12cm]{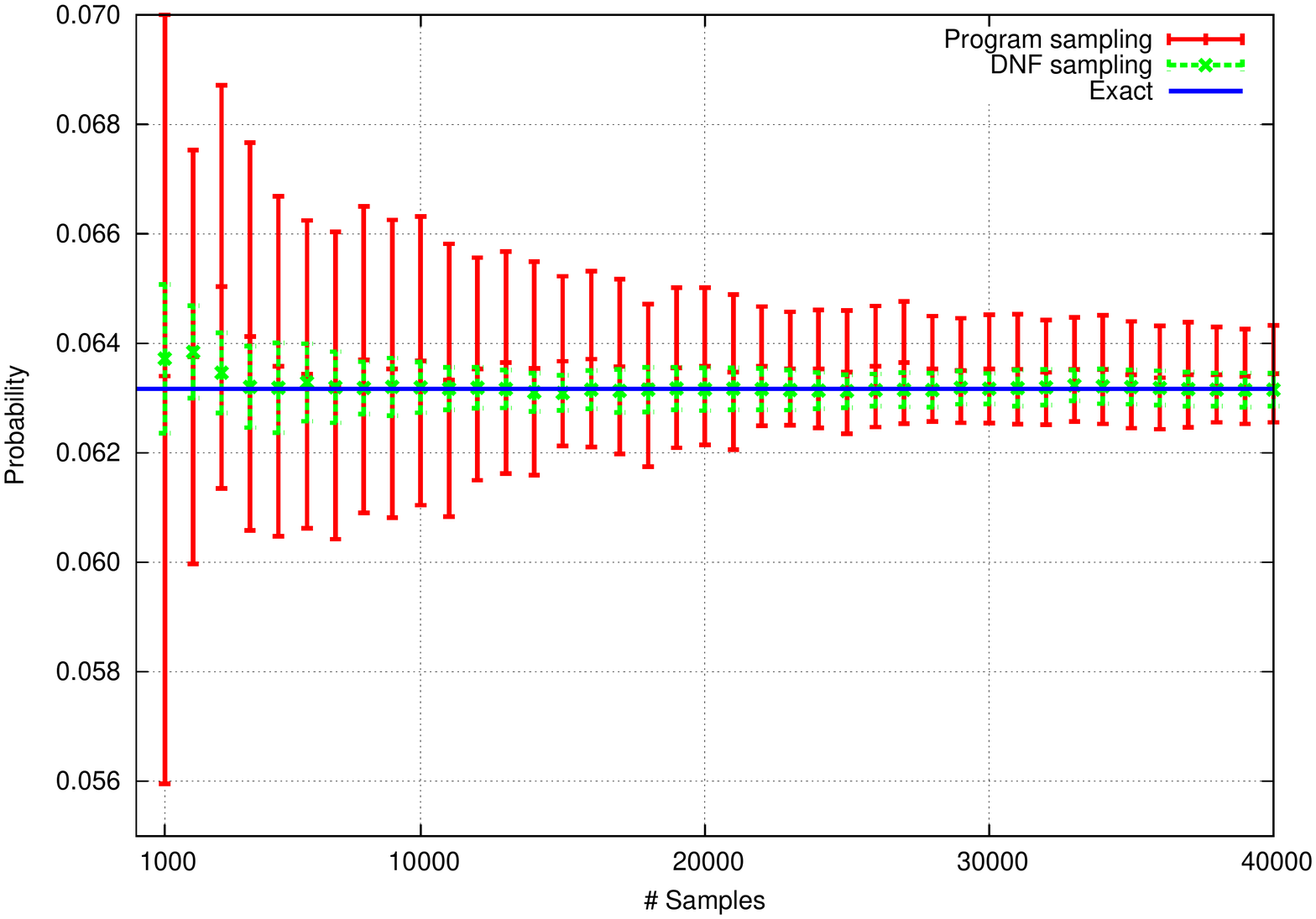}
  \caption{DNF Sampling convergence vs. Program Sampling convergence vs. Exact, for: \texttt{path('HGNC\_582','HGNC\_983',7)}.}
  \label{fig:alzw3_9}
\end{figure}

Finally, we present in Figure~\ref{fig:alzw3_9} the last benchmark. The query used for that experiment is actually a query containing many failing derivations, because of that Program Sampling has difficulty at converging. Here we can see significantly better convergence by DNF Sampling.

Our experiments have shown that DNF Sampling is not as potent as we expected compared with Program Sampling, still we saw immediate convergence for problems where the proofs are mutually exclusive and we have indications that for problems where the proofs will have some exclusiveness DNF Sampling is very potent.

\section{Related Work}
As mentioned earlier, DNF Sampling is based on the general sampling scheme introduced in~\cite{KARP83}. This scheme has been used for probability estimation in the context of probabilistic databases~\cite{KEIJZER04,Re07} and probabilistic graph mining~\cite{Zou09}. In the context of statistical relational learning, the scheme has been used to estimate the number of true groundings of a clause~\cite{1102407}. While the use of sampling in combination with a reduction to a DNF formula for probabilistic logic programs has already been proposed (but not realized) by~\cite{Dantsin91}, to the best of our knowledge, this paper is the first to actually use DNF Sampling for inference in a probabilistic logic programming system. The ProbLog system also includes approximate inference methods that do not use sampling, but rely on restricting the number of proofs encoded in the DNF~\cite{KIMMIG08}.

\section{Conclusions and Future Work}
We have introduced DNF Sampling, a new sampling method for approximate inference in ProbLog. DNF Sampling exploits the same reduction to DNF as ProbLog's exact inference method. It is a valuable addition to the ProbLog system, as it allows one to exploit the already constructed DNF even if its exact processing turns out to be infeasible. However, our experimental comparison of the new method with Program Sampling, ProbLog's previous sampling based inference technique, indicates that Program Sampling outperforms DNF Sampling for approximate inference in biological networks, and should thus be preferred if it is expected that exact inference will be infeasible.
Future work includes the application of DNF Sampling in the context of the nested tries structure used in tabled ProbLog~\cite{231065}, where solving the disjoint-sum-problem becomes intractable while a compact representation of the underlying formula is available. As this representation exploits the presence of shared structure in the formula, extending DNF Sampling to this case is a promising approach to further increase the scalability of ProbLog inference.

\section*{Acknowledgements}
This research is supported by GOA/08/008 ``Probabilistic Logic Learning". Angelika Kimmig is supported by the Research Foundation-Flanders (FWO-Vlaanderen).

\bibliographystyle{splncs}
\bibliography{CICLOPS2010}

\end{document}